\documentclass[epj]{webofc}
\usepackage[varg]{txfonts}   
\woctitle{MENU 2013}
\def \ee {e^+e^-}
\begin{document}
\begin{flushright}
{\sf  MITP/14-013 \\
  } 
\end{flushright}
\vspace{-0.5cm}
\title{Pseudoscalar Transition Form Factors from Rational Approximants}
%
%

\author{Pere Masjuan\inst{1}\fnsep\thanks{\email{masjuan@kph.uni-mainz.de}. Supported by the Deutsche Forschungsgemeinschaft DFG through the Collaborative Research Center ``The Low-Energy Frontier of the Standard Model" (SFB 1044) and by the Jefferson Science Associates, LLC.}
}

\institute{PRISMA Cluster of Excellence, Institut f\"ur Kernphysik, Johannes Gutenberg-Universit\"at, D-55099 Mainz, Germany}

\abstract{%
The $\pi^0$, $\eta$, and $\eta^\prime$ transition form factors in the space-like region are analyzed at low and intermediate energies in a model-independent way through the use of rational approximants.  Slope and curvature parameters as well as their values at infinity are extracted from experimental data. These results are suited for constraining hadronic models such as the ones used for the hadronic light-by-light scattering piece of the anomalous magnetic moment of the muon, and for  the mixing parameters of the $\eta - \eta^\prime$ system.
}
\maketitle
\vspace{-0.6cm}
\section{Introduction}
\label{intro}

The pseudoscalar transition form factor (TFF) $\gamma^*\gamma^* - P$, where $P=\pi^0, \eta, \eta^\prime\cdots$, have acquired a lot of attention recently, both from the experimental and theoretical sides,  since the release of the \textit{BABAR} data on the $\pi^0$-TFF in 2009 \cite{Aubert:2009mc}. The TFF describes the effect of the strong interaction on the $\gamma^*\gamma^* - P$ transition and is represented by a function $F_{P\gamma^*\gamma^*}(q_1^2,q_2^2)$ of the photon virtualities $q_1^2$, and $q_2^2$.

From the experimental point of view, one can study the TFF from both space-like and time-like energy regimes. The time-like TFF can be accessed from a single Dalitz decay $P \to l^+l^- \gamma$ process which contains an off-shell photon with the momentum transfer $q_1^2$ and defines a $F_{P\gamma^*\gamma}(q_1^2,0)$ covering the $4m_l^2<q^2<m_P^2$ region.  The space-like TFF can be accessed in $\ee$ colliders by the two-photon-fusion reaction $e^+e^-\to e^+e^-P$. The common practice is to extract the TFF when one of the outgoing leptons is tagged and the other is not, that is, the single-tag method.
The tagged lepton emits a highly off-shell photon with the momentum transfer $q_1^2\equiv -Q^2$ and is detected,
while the other, untagged, is scattered at a small angle and its momentum transfer $q_2^2$ is near zero, i.e., 
$F_{P\gamma^*\gamma}(Q^2)\equiv F_{P\gamma^*\gamma^*}(-Q^2,0)$.

Theoretically, the limits $Q^2=0$ and $Q^2\rightarrow\infty$ are well known in terms of the axial anomaly in the chiral limit of QCD \cite{Adler:1969gk} and pQCD \cite{Lepage:1980fj}, respectively. The TFF is then calculated as a convolution of a perturbative hard-scattering amplitude and a gauge-invariant meson distribution amplitude (DA)~\cite{Mueller:1994cn} which incorporates the nonperturbative dynamics of the QCD bound-state~\cite{Lepage:1980fj}. Some model needs to be used either for the DA or the TFF itself~\cite{Lepage:1980fj}. The discrepancy among different approaches reflects the model-dependency of that procedure. A different procedure might be, then, desirable.

\section{Transition form factors from rational approximants}

We propose~\cite{Masjuan:2012wy,EscribanoMasjuan}  to use a sequence of rational approximants $P^N_M=\frac{{\cal P}_N}{{\cal P}_M}$ called Pad\'e approximants (PA)~\cite{Baker} constructed from the Taylor expansion of the $F_{P\gamma^*\gamma}(Q^2)$ to fit the space-like data~\cite{Aubert:2009mc,SL,Uehara:2012ag,Acciarri:1997yx} and obtain, in such a way, the derivatives of the $F_{P\gamma^*\gamma}(Q^2)$ at the origin of energies in a simple, systematic and model-independent way~\cite{Masjuan:2012wy,EscribanoMasjuan}. 

Since the analytic properties of TFFs are not known, the kind of PA sequence to be used is not determine in advance. We consider two different sequences and the comparison among them should reassess our results. The first one is a $P^L_1(Q^2)$ sequence inspired by the success of the simple vector meson dominance ansatz~\cite{Masjuan:2012wy}, and the second one is a $P^N_N(Q^2)$ sequence which satisfy the pQCD constrains $Q^2 F_{P\gamma \gamma^*}(Q^2) \sim \textrm{constant} $. After combining both sequence's results, slope and curvature results are shown in Table~\ref{tab1}, where $\lim_{Q^2\to\infty }Q^2F_{P\gamma^*\gamma}(Q^2) $ from the $P^N_N(Q^2)$ is also shown.

\begin{figure}
\centering
\includegraphics[width=7.0cm,clip]{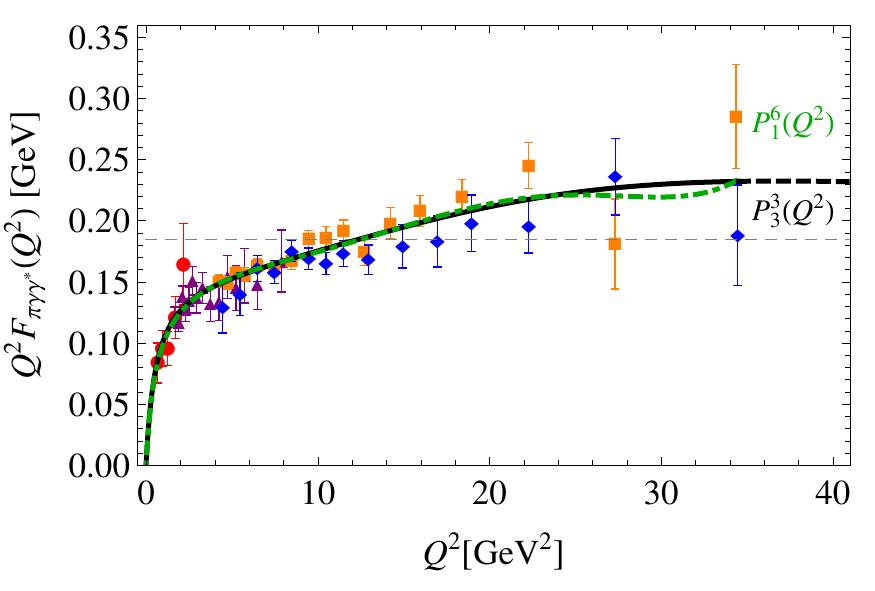}
\includegraphics[width=7.0cm,clip]{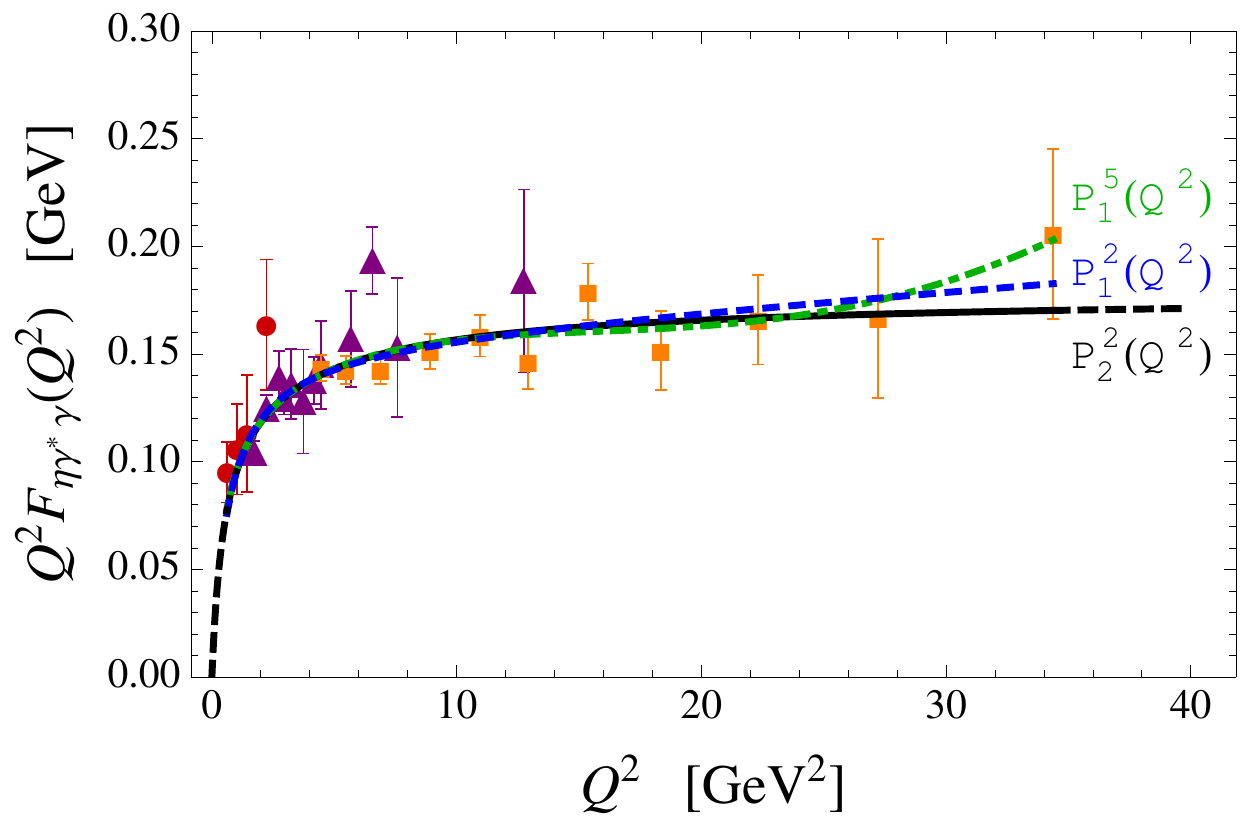}
\includegraphics[width=7.0cm,clip]{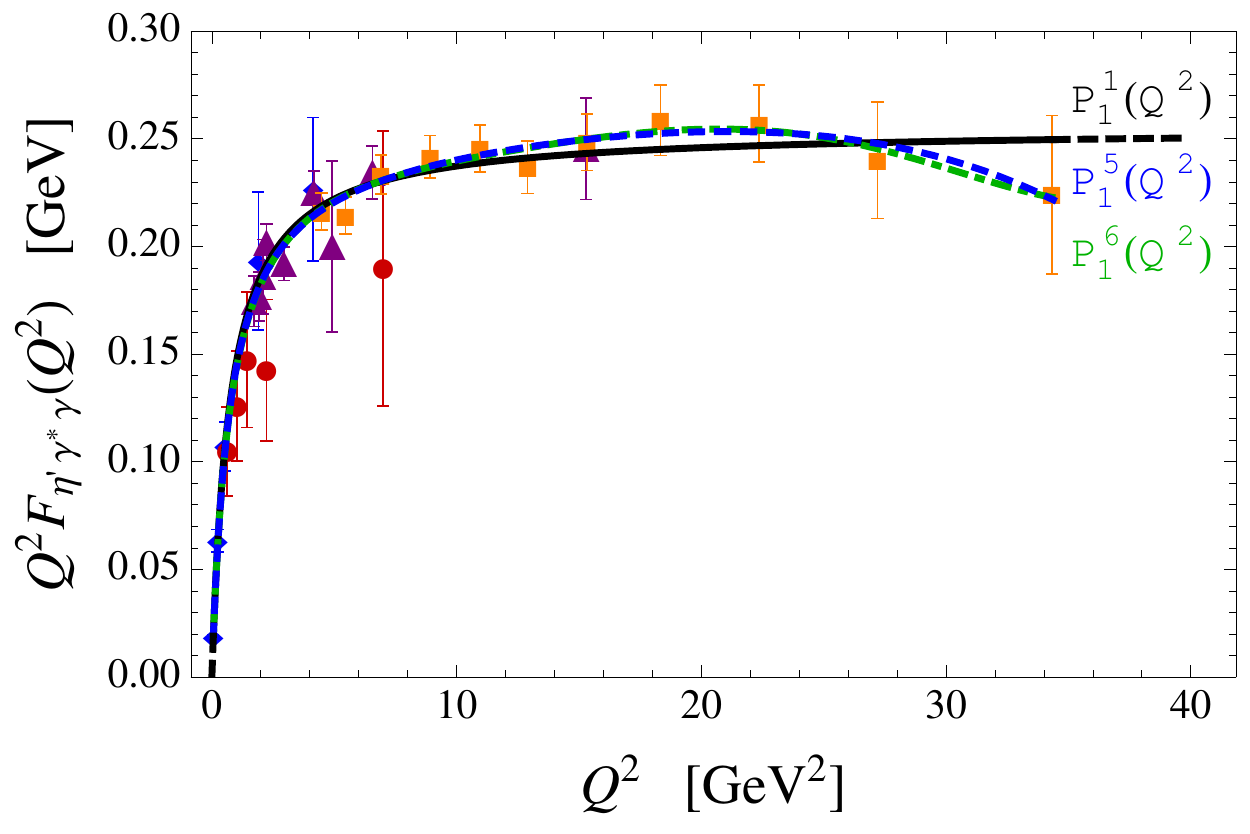}
\caption{$\pi^0$ (left upper panel), $\eta$ (right upper panel), and $\eta^\prime$ (lower panel) TFFs. 
Green-dot-dashed lines show our best $P^L_{1}(Q^2)$ fit, and black-solid lines show our best $P^N_{N}(Q^2)$ fit. Black-dashed lines display the extrapolation of the $P^N_{N}(Q^2)$ at $Q^2=0$ and $Q^2\to\infty$. Experimental data are from CELLO (red circles), CLEO (purple triangles), and {\textit{BABAR}} (orange squares) Colls.~\cite{SL}. The $\pi^0$ figure contains also data from BELLE (blue diamonds)~\cite{Uehara:2012ag}; and the $\eta^\prime$ figure data from L3 (blue diamonds) \cite{Acciarri:1997yx}.}
\label{fig1}       
\end{figure}

\begin{table}
\centering
\caption{$\pi^0,\eta$, and $\eta^\prime$ slope $b_P$, curvature $c_P$, asymptotic limit, and contribution to HLBL.}
\label{tab1}       
\begin{tabular}{ccccc}
\hline
 & $b_P$ & $c_P$ & $\lim_{Q^2\to\infty }Q^2F_{P\gamma^*\gamma}(Q^2) $  & $a_{\mu}^{\textrm{HLBL;P}}$  \\\hline
$\pi^0$ & $0.0324(22)$ & $1.06(27)\cdot 10^{-3}$ & $2 f_{\pi} $ & $6.49(56) \cdot 10^{-10}$  \\
$\eta$ & $0.60(7)$ & $0.37(12)$ & $0.160(24) \textrm{GeV}$ & $1.25(15) \cdot 10^{-10}$\\ 
$\eta^\prime$ & $1.30(17)$ & $1.72(58)$ & $0.255(4) \textrm{GeV}$  & $1.27(19) \cdot 10^{-10}$\\ \hline
\end{tabular}
\end{table}

The low-energy parameters obtain with this method can be used to constrain hadronic models with resonances used to account for the hadronic light-by-light scattering contribution part (HLBL) of the anomalous magnetic moment of the muon~\cite{g2}, rare pseudoscalar decays and continuum cross section determinations in the charmonium region~\cite{inprep}.

\section{Applications of the transition form factors}
\subsection{Hadronic light-by-light scattering contribution to the muon $(g-2)$}

The HLBL cannot be directly related to any measurable cross section and requires the knowledge of QCD contributions at all energy scales. Since this is not known yet, one needs to rely on hadronic models to compute it. Such models introduce some systematic errors which are difficult to quantify. The large-$N_c$ limit of QCD~\cite{largeNc} provides a very useful framework to approach this problem but has, however, a shortcoming. Calculations carried out in the large-$N_c$ limit demand an infinite set of resonances. As such sum is not known in practice, one ends up truncating the spectral function in a resonance saturation scheme, the so-called Minimal Hadronic Approximation \cite{Peris:1998nj}. The resonance masses used in each calculation are then taken as the physical ones from PDG \cite{PDG2012} instead of the corresponding (but unknown) masses in the large-$N_c$ limit. Both problems might lead to large systematic errors not included so far \cite{Masjuan:2007ay,Masjuan:2012wy,Masjuan:2012gc}.

It was pointed out in Ref.~\cite{Masjuan:2007ay} that, in the large-$N_c$ framework, the Minimal Hadronic Approximation can be understood from the mathematical theory of Pad\'e Approximants (PA) to meromorphic functions. Obeying the rules from this mathematical framework, one can compute the desired quantities in a model-independent way and even be able to ascribe a systematic error to the approach \cite{Masjuan:2009wy}. One interesting detail from this theory \cite{Baker} is that given a low-energy expansion of a meromorphic function, a PA sequence converges much faster than a rational function with the poles fixed in advance (such as the common hadronic models used so far for evaluating the HLBL), especially when the correct large $Q^2$ behavior is imposed.

We will use, instead of a hadronic model for the TFF, a sequence of PA~\cite{Masjuan:2007ay} build up from the low-energy expansion obtained in the previous section. The TFF is considered to be off-shell. To match the large momentum behavior with short-distance constraints from QCD, calculable using the OPE, we consider the relations obtained in Ref.~\cite{g2}. In practice this amounts to use for the TFFs the expression: 
\begin{equation}\label{P01}
\begin{split}
&F_{P\gamma^*\gamma^*}^{P^0_1}(p_{P}^2,q_1^2,q_2^2)\, =\, a \frac{b}{q_1^2-b}\frac{b}{q_2^2-b}\big(1+c\, p_{P}^2\big)\, ,
\end{split}
\end{equation}
\noindent
where $p_{P}=q_1+q_2$ and the free parameters are matched at low energies \cite{Masjuan:2012wy}: $a$ is fixed by $\Gamma_{P\rightarrow \gamma\gamma}$ from PDG~\cite{PDG2012}; and $b$ by a matching to the slope $b_{P}$ from Table~\ref{tab1}. The parameter $c$ characterizes the off-shellnes of the pseudoscalar and is obtained by imposing, along the lines of the Pad\'e method, that $\lim_{q\rightarrow \infty} F_{\pi^*\gamma^*\gamma^*}^{P01}(q^2,q^2,0)\, = f_{\pi} \chi /3, $where $\chi =( -3.3 \pm 1.1)$ GeV$^{-2}$, with an error of $30\%$ as proposed in Ref.~\cite{g2}, and $c=0$ for $\eta$ and $\eta^\prime$. Our results are collected in Table~\ref{tab1} and are in nice agreement with most of the recent determinations~\cite{g2,Roig:2014uja}.

\subsection{$\eta-\eta^\prime$ mixing parameters}

The physical $\eta$ and $\eta'$ mesons are an admixture of the $SU(3)$ Lagrangian eignestates~\cite{Leutwyler:1997yr}. Deriving the parameters governing the mixing is a challenging task. Usually, these are determined through the use of $\eta(')\rightarrow \gamma\gamma$ decays as well as vector radiative decays into $\eta(')$ together with $\Gamma(J/\Psi \to \eta^\prime \gamma)/\Gamma(J/\Psi \to \eta \gamma)$~\cite{Leutwyler:1997yr}. However, since pQCD predicts 
that the asymptotic limit of the TFF for the $\eta(')$ is essentially given in terms of these mixing 
parameters, we use our TFF parametrization to estimate the asymptotic limit and further constrain the mixing parameters with compatible results compared to standard (but more sophisticated) determinations.

To determine the mixing parameters~\cite{EscribanoMasjuan} we use the normalization at zero of both TFFs  from the measured decay widths $\Gamma_{\eta\to\gamma\gamma}=0.516(18)$ keV and $\Gamma_{\eta^\prime\to\gamma\gamma}=4.35(14)$ keV, respectively, and the asymptotic value of the $\eta$ TFF $\lim_{Q^2\to\infty }Q^2F_{\eta\gamma^*\gamma}(Q^2)=0.160(24)$ GeV~\cite{EscribanoMasjuan}:
\begin{equation}
\label{eq:mixresults}
F_q/F_\pi=1.06(1)\ ,\quad
F_s/F_\pi=1.56(24)\ , \quad \phi=40.3(1.8)^{\circ}\ ,
\end{equation}
with $F_\pi=92.21(14)$ MeV \cite{PDG2012}.
They can be compared, for instance, with the determination of the mixing parameters obtained in Ref.~\cite{Leutwyler:1997yr},
after a careful analysis of $V\to\eta^{(\prime)}\gamma$, $\eta^{(\prime)}\to V\gamma$, with $V=\rho, \omega, \phi$, and
$\eta^{(\prime)}\to\gamma\gamma$ decays, and the ratio $R_{J/\psi}\equiv\Gamma(J/\psi\to\eta^\prime\gamma)/\Gamma(J/\psi\to\eta\gamma)$, with the latest experimental measurements of these decays gives
$F_q/F_\pi=1.07(1)$, $F_s/F_\pi=1.63(3)$ and $\phi=39.6(0.4)^{\circ}$.
The agreement between these determinations and the values in Eq.~(\ref{eq:mixresults}) is quite impressive
since we only use the information of the TFFs to predict the mixing parameters.


\begin{thebibliography}{99}
\setlength{\itemsep}{0cm}

\bibitem{Aubert:2009mc}
  B.~Aubert {\it et al.}  [BaBar Collaboration],
  Phys.\ Rev.\ D {\bf 80} (2009) 052002.

\bibitem{Adler:1969gk}
  S.~L.~Adler,
  Phys.\ Rev.\  {\bf 177} (1969) 2426;
  J.~S.~Bell and R.~Jackiw,
  Nuovo Cim.\ A {\bf 60} (1969) 47.

  
\bibitem{Lepage:1980fj}
  G.~P.~Lepage and S.~J.~Brodsky,
  Phys.\ Rev.\ D {\bf 22} (1980) 2157;
  T.~Feldmann and P.~Kroll,
 Phys.\ Rev.\ D {\bf 58} (1998) 057501
 [hep-ph/9805294].

  
\bibitem{Mueller:1994cn}
  D.~Mueller,
  Phys.\ Rev.\ D {\bf 51} (1995) 3855.

  
  
\bibitem{Masjuan:2012wy}
  P.~Masjuan, S.~Peris and J.~J.~Sanz-Cillero, 
  Phys.\ Rev.\ D {\bf 78} (2008) 074028; 
    P.~Masjuan,
  Phys.\ Rev.\ D {\bf 86} (2012) 094021.
  
  \bibitem{EscribanoMasjuan}
  R.~Escribano, P.~Masjuan and P.~Sanchez-Puertas,
  Phys.\ Rev.\ D {\bf 89} (2014) 034014
  [hep-ph/1307.2061].
  
  
 \bibitem{Baker}
  G.A.Baker and P. Graves-Morris,
  Encyclopedia of Mathematics and its Applications, Cambridge Univ. Press, 1996; 
  P.~Masjuan Queralt,
  arXiv:1005.5683 [hep-ph].
  
  \bibitem{SL}
  H.~J.~Behrend {\it et al.}  [CELLO],
  Z.\ Phys.\ C {\bf 49} (1991) 401;
  J.~Gronberg {\it et al.}  [CLEO],
  Phys.\ Rev.\ D {\bf 57} (1998) 33;
  P.~del Amo Sanchez {\it et al.}  [BaBar],
  Phys.\ Rev.\ D {\bf 84} (2011) 052001.
  
  \bibitem{Uehara:2012ag}
  S.~Uehara {\it et al.}  [Belle Coll.],
  Phys.\ Rev.\ D {\bf 86} (2012) 092007;

\bibitem{Acciarri:1997yx}
  M.~Acciarri {\it et al.}  [L3 Coll.],
  Phys.\ Lett.\ B {\bf 418} (1998) 399.

  
\bibitem{g2}    
  F.~Jegerlehner and A.~Nyffeler,
  Phys.\ Rept.\  {\bf 477} (2009) 1;
   A.~Nyffeler,
  Phys.\ Rev.\ D {\bf 79} (2009) 073012; 
    J.~Prades, E.~de Rafael and A.~Vainshtein,
  [arXiv:0901.0306 [hep-ph]];
  P.~Masjuan and M.~Vanderhaeghen,
  arXiv:1212.0357 [hep-ph].

\bibitem{inprep}
P. Masjuan and P. Sanchez-Puertas, in preparation.  
   
\bibitem{largeNc}
  G.~'t Hooft,
  Nucl.\ Phys.\ B {\bf 72} (1974) 461;
    E.~Witten,
  Nucl.\ Phys.\ B {\bf 160} (1979) 57.

\bibitem{Peris:1998nj}
  S.~Peris, M.~Perrottet and E.~de Rafael,
  JHEP {\bf 9805} (1998) 011.

\bibitem{PDG2012}
  J.~Beringer {\it et al.}  [Particle Data Group Collaboration],
  Phys.\ Rev.\ D {\bf 86} (2012) 010001.

\bibitem{Masjuan:2007ay}
  P.~Masjuan and S.~Peris,
  JHEP {\bf 0705} (2007) 040;
  Phys.\ Lett.\ B {\bf 663} (2008) 61.
  

\bibitem{Masjuan:2012gc}
  P.~Masjuan, E.~Ruiz Arriola and W.~Broniowski,
  Phys.\ Rev.\ D {\bf 85} (2012) 094006;
  Phys.\ Rev.\ D {\bf 87} (2013) 014005.
  
\bibitem{Masjuan:2009wy}
  P.~Masjuan and S.~Peris,
  Phys.\ Lett.\ B {\bf 686} (2010) 307;
    P.~Masjuan and J.~J.~Sanz-Cillero,
  Eur.\ Phys.\ J.\ C {\bf 73} (2013) 2594.


\bibitem{Leutwyler:1997yr}
  H.~Leutwyler,
  Nucl.\ Phys.\ Proc.\ Suppl.\  {\bf 64} (1998) 223;
  T.~Feldmann, P.~Kroll and B.~Stech,
  Phys.\ Rev.\ D {\bf 58} (1998) 114006;
  R.~Escribano and J.~-M.~Frere,
  JHEP {\bf 0506} (2005) 029.

  
  \bibitem{Roig:2014uja}
  P.~Roig, A.~Guevara and G.~L\'op.~Castro,
  arXiv:1401.4099 [hep-ph].
    
 \end{thebibliography}
\end{document}